# ContractGuard: Defend Ethereum Smart Contracts with Embedded Intrusion Detection

Xinming Wang, Jiahao He, Zhijian Xie, Gansen Zhao, Shing-Chi Cheung

**Abstract**— Ethereum smart contracts are programs that can be collectively executed by a network of mutually untrusted nodes. Smart contracts handle and transfer assets of values, offering strong incentives for malicious attacks. Intrusion attacks are a popular type of malicious attacks. In this paper, we propose *ContractGuard*, the first intrusion detection system (IDS) to defend Ethereum smart contracts against such attacks. Like IDSs for conventional programs, ContractGuard detects intrusion attempts as abnormal control flow. However, existing IDS techniques/tools are inapplicable to Ethereum smart contracts due to Ethereum's decentralized nature and its highly restrictive execution environment. To address these issues, we design ContractGuard by embedding it in the contracts to profile context-tagged acyclic paths, and optimizing it under the Ethereum gas-oriented performance model. The main goal is to minimize the overheads, to which the users will be extremely sensitive since the cost needs to be paid upfront in digital concurrency. Empirical investigation using real-life contracts deployed in the Ethereum mainnet shows that on average, ContractGuard only adds to 36.14% of the deployment overhead and 28.27% of the runtime overhead. Furthermore, we conducted controlled experiments and show that ContractGuard successfully guard against attacks on all real-world vulnerabilities and 83% of the seeded vulnerabilities.

**Index Terms**—Blockchain, Ethereum Smart Contract, Intrusion Detection System, Anomaly Detection

✦

## 1 INTRODUCTION

Blockchains as one form of distributed ledger technology have gained considerable interest and adoption since Bitcoin [1] was introduced. Participants in a blockchain system collectively run a consensus protocol to maintain and secure a shared ledger of data. Blockchains were initially introduced for peer-to-peer payments, but more recently, it has been extended to allow programmable transactions in the form of *smart contracts*. Smart contracts are programs that can be collectively executed by a network of mutually untrusted nodes, which implement a consensus protocol that digitally enforce agreements among nodes: neither the nodes nor the creator of the smart contract can feasibly modify its code or subvert its execution.

Because smart contracts are entrusted by the users to handle and transfer assets of considerable value, they are subject to intensive hacking activities. Such hacking is more dangerous than that on a conventional network system, because once deployed on the blockchain, the contract becomes immutable, essentially creating a high-risk paradigm: the deployed code is nearly impossible to patch, and contracts collectively control billions of USD worth of digital assets. There have been a plenty of well-documented attacks on the Ethereum smart contracts [3].

In the literature, considerable research efforts have been spent on static analysis tools that detect vulnerabilities *before deployment* [4]–[11]. However, to the best of our knowledge, there is currently no security tools to protect smart contracts *after deployment* -- once being deployed on Ethereum, smart contract programs are completely exposed to the intruders. If the attack is accepted into the main chain, the loss will be irreversible. *There is no second chance to fix the mistake.*

For conventional systems, administrators recognize that despite the developers' best efforts to fix defects, vulnerabilities are prevalent. *Intrusion detection systems* (IDS) [12] are commonly applied as a major means of protecting deployed systems from security attacks. By the way of how they detect intrusions, IDSs can be classified as signature-based or anomaly-based [12]. A typical anomaly-based IDS monitors dynamic program behavior against normal program behavior and raises an alert when detecting an anomaly. The normal behavior is learnt through training and profiling.

In this work, our goal is to design a practical anomaly-based IDS that can protect smart contracts after deployment. Fig. 1 illustrates the basic idea. When detecting abnormal behavior, the IDS will rollback all the changes to the contract states and raise an alarm to the administrators. This provides a chance to prevent the irreparable loss that the vulnerability can cause. For example, had the administrator of the DAO [13] applied IDS to protect their smart contracts, the famous DAO attack would have been prevented, because the attack triggered an abnormal reentrant control flow not observed in in-house testing. Similarly, the Parity Multisig Wallet incidence [14] could also have been prevented, because the attack triggered a call of library code with an abnormal calling-context. Essentially, such an IDS can provide a second chance for the administrators to save the catastrophe.

To put such an idea into practical work, however, we have to consider the unique characteristics of blockchain in general and Ethereum in particular. These characteristics make conventional IDS techniques inapplicable to smart contracts:

- X.M. Wang is with the Lakala Group, Building D1, Zhongguancun Yihao, Beiqing Road, Beijing, China. E-mail: wangxinming@lakala.com
- J.H. He, Z.J. Xie, G.S. Zhao are with the School of Computer Science, South China Normal University, Zhongshan Road No. 55, Guangzhou, Guangdong, China. E-mail:{hejiahao, xiezhijian, gzhao}@m.scnu.edu.cn.
- S.C. Cheung is with the Department of Computer Science and Engineering, The Hong Kong University of Science and Technology, Clear Water Bay, Kowloon, Hong Kong. E-mail: scc@cse.ust.hk
- Corresponding authors: Gansen Zhao; S.C. Cheung;

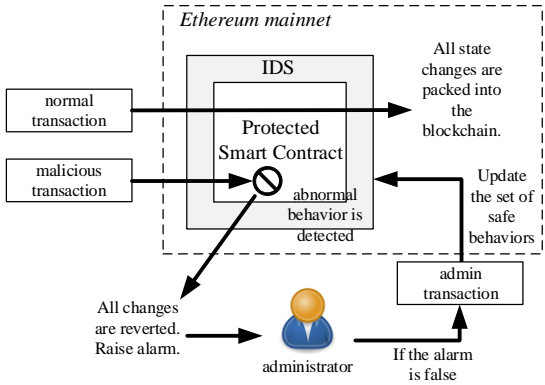

Fig. 1. The idea of IDS for smart contracts.

(1) As Ethereum is essentially a *decentralized application platform* [2], the smart contract programs are executed by mutually untrusted nodes across the whole world. Therefore, neither the conventional host-based nor network-based IDS techniques [12] are applicable. Instead, we must embed the IDS into the contracts, so that the protection procedure becomes part of the consensus and none of the nodes can bypass them.

(2) Ethereum smart contracts run in a highly restrictive environment called *Ethereum virtual machine* (EVM), which lacks many capabilities that facilitate the implementation of conventional IDSs, such as hardware registers [15], call stack traversal [16], and event hooks [17]. Furthermore, EVM imposes a strict size limitation of 24576 bytes on every contract. This requires a very careful design to minimize the size of embedded IDS implementation.

(3) Ethereum introduces a fundamentally different performance model. While the overheads of conventional IDSs are optimized toward reducing execution time, for Ethereum smart contracts the execution time is irrelevant. Instead, the overheads are determined by the *gas cost* [18]. This requires gas-oriented overhead optimization that is different from those used in conventional IDSs.

In this paper, we propose *ContractGuard* as the first anomaly-based IDS to address the above challenges for Ethereum smart contracts. Compared with the signature-based approach, the anomaly-based approach has the advantage of not requiring the signatures of known vulnerabilities. Following many existing anomaly-based IDSs [15], [19], [20], ContractGuard monitors control flow path to classify execution behavior. The key idea is to combine the following two techniques to meet the stringent requirements on the effectiveness and efficiency of IDS for smart contracts:

(1) **Context-tagged acyclic path profiling**: ContractGuard uses the classic algorithm proposed by Ball and Larus [21] to index and profile intraprocedural acyclic path segments efficiently. Furthermore, to improve the effectiveness of anomaly detection, ContractGuard adds calling-context information to each intraprocedural path. As EVM does not provide the capability to traverse call stack directly, a profiling scheme is introduced to instrument calling-contexts.

(2) **Gas-efficient adaptive path set storage**: As EVM heavily penalizes the use of persistent contract storage, conventional memory-based set implementations are highly gas-inefficient. Instead, ContractGuard adaptively selects one of the three data structures (embedded list, embedded minimal perfect hash table [22], and built-in mapping). As a result, ContractGuard is able to achieve practical runtime and deployment gas overhead.

Note that ContractGuard implements all the above techniques on EVM binary code. Therefore, administrators can use it to protect Ethereum smart contracts without the need of their Solidity source code.

To validate our approach, we conduct three sets of experiments using real world Ethereum smart contracts. The first set of experiments concerns whether ContractGuard is practical in terms of gas overhead and false alarms. We applied ContractGuard on the 8,314 contracts deployed on Ethereum mainnet up to block number 4,200,000, or the date Aug-24-2017, with at least 100 transactions. The result show that the average deployment and runtime overhead of ContractGuard are 36.14% and 28.27%, respectively. The second set of experiment investigates the effectiveness of ContractGuard by applying it on six real-life smart contract programs with reported vulnerabilities. The result shows that ContractGuard successfully defend all of them. The third set of experiments concerns the effectiveness using seeded vulnerabilities and real-life developer test cases for realistic training. ContractGuard detect 83% of the seeded vulnerabilities.

From a technical standpoint, this paper makes the following contributions:

1. We present a detailed study on whether attacks on smart contract vulnerabilities can be defended by detecting abnormal control flow paths.
2. We present the first empirical evidence that deploying IDS for Ethereum smart contracts is practical in term of deployment/runtime overhead and false alarms, using real world data from Ethereum mainnet.
3. We propose ContractGuard as the first anomaly-based IDS for Ethereum smart contracts.
4. We demonstrate that ContractGuard scales to real life smart contract programs and it effectively defends both real-life and seeded vulnerabilities.

The rest of this paper is organized as follows. Section 2 introduces the background on the smart contract program model and the context-tagged acyclic path. Section 3 gives details and insights into motivation of ContractGuard. Section 4 presents the design of ContractGuard. Section 5 reports the result of simulation experiments, Section 6 reports the experiment results. After discussing the related work (Section 7), we close with the conclusions (Section 8).

## 2 BACKGROUND

### 2.1 Ethereum Smart Contract Program Model

As shown in Fig. 2, an Ethereum smart contract program is composed of one or more *contract accounts* that are deployed instances of contract code, and one or more *user accounts* that hold *ethers* as the cryptocurrency. The users rely on *clients* to interact with the Ethereum network, which is maintained by a group of *miners* that pack transactions into

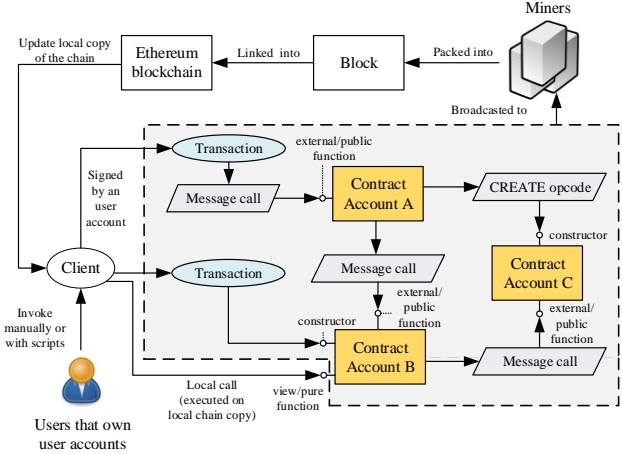

Fig. 2. Ethereum Smart contract program model.

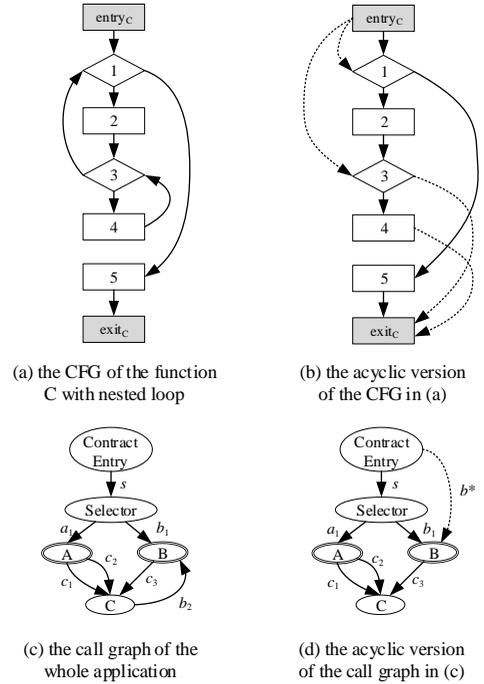

(a) the CFG of the function C with nested loop

(b) the acyclic version of the CFG in (a)

(c) the call graph of the whole application

(d) the acyclic version of the call graph in (c)

Fig. 3. Examples of Context-tagged Acyclic Path.

the blockchain.

**Program structure**: Smart contracts are mostly written in Solidity [23], which is a JavaScript-like, but typed, programming language. At the source level, contracts written in Solidity are analogous to classes in object-oriented languages. Contracts can contain declarations of state variables, definitions of functions, modifiers, and constructors. Functions support encapsulation (visibility attributes). To enable deployment on the Ethereum platform, developers compile the contract source code into EVM binary code. A piece of code called *function selector* is added as an entry point of the contract. Whenever a function is called, the contract starts executing at the function selector. The selector decodes the message and jumps into an appropriate function. Ethereum smart contracts allow customizing the handling of messages that do not specify a concrete calling function through the *fallback function*.

**Invocation model**: After deployment, contract accounts can be invoked with its external/public functions in three different ways. The first way is by the client sending a *message call transaction*, which makes a message call that contains the signature hash of the target function and parameter data. A message call transaction must be signed, broadcasted, and packed to take effect. It is a write-operation that can change the account balances and the contract states. Miners charge the sender with ethers to pay for the gas costed by the transaction. The second way is by another contract account directly making a message call to invoke a function. This action can always trace transitively back to a message call transaction. Finally, the third way is by the client locally calling view/pure functions, which never modify the state and does not cost gas.

**Data management model**: The EVM is a 256-bit word stack-based machine. Contracts can access four kinds of data: the operation *stack* with a maximum size of 1024, a word-addressed byte array called *memory* to serve as volatile storage, a word-addressed word array called *storage* to serve as persistent storage, and a read-only word array *calldata* as a copy of message call input data. The contents of stack, memory, and calldata are not persisted at the end of transactions, nor are they accessible across the boundary of message calls.

## 2.2 Context-tagged Acyclic Path

ContractGuard protects smart contracts by monitoring context-tagged acyclic path, which is extended from the intra-procedural acyclic path defined by Ball and Laurus [21]. A context-tagged acyclic path is a tuple $\langle p, c \rangle$, where $p$ is an intra-procedural acyclic path and $c$ is the calling-context. Informally, intra-procedural acyclic path is a path on the acyclic version of control flow graph (CFG). This acyclic version is obtained by replacing every backedge $w \to v$ with two surrogate edges $entry \to v$ and $w \to exit$. For example, consider the CFG in Fig. 3 (a). The two backedges $3 \to 1$ and $4 \to 3$ are replaced by surrogate edges $entry_c \to 3$, $entry_c \to 4$, $3 \to exit_c$ and $4 \to exit_c$ in Fig. 3 (b).

Intra-procedural acyclic path has two interesting properties that make it particularly suitable to the task of anomaly detection. Firstly, the numbers of intra-procedural acyclic paths are always limited, as the acyclic CFG does not have cycles. This is important to the efficient storage of path set. Secondly, a full path in the original CFG can be broken into several acyclic paths at the backedges. For example, the path $entry_c \to 1 \to 2 \to 3 \to 4 \to 3 \to 4 \to 3 \to 4 \to 3 \to 1 \to 5 \to exit_c$ in Fig. 3 (a) can be broken into $p_1$: $entry_c \to 1 \to 2 \to 3 \to 4 \to exit_c$, $p_2$: $entry_c \to 3 \to 4 \to exit_c$, $p_3$: $entry_c \to 3 \to 4 \to exit_c$, $p_4$: $entry_c \to 3 \to exit_c$, and $p_5$: $entry_c \to 1 \to 5 \to exit_c$. Intuitively, acyclic paths correspond to the path segments before, inside or after a loop. Full paths with different numbers of loop iterations can share the same set of acyclic paths. The loop-insensitivity can be important to the reduction of the false alarms.

Given the call graph with functions as nodes and callsites as edges, the calling context of a function $f$ is defined as the sequence of call sites pending on entry to $f$ [24]. Due to recursion, the call graph may contain cycles. In the same way as a backedge is replaced by a subrogate edge in a CFG, each recursive callsite is replaced by a subrogate callsite from the program entry to the function that is called recursively. For example, the call graph of the contract in Fig. 3 (c) is transformed to that in Fig. 3 (d). As a result, there are four calling-contexts of C: $s \to a_1 \to c_1$, $s \to a_1 \to c_2$, $s \to b_1 \to c_3$, and $b^* \to c_3$.

In order to capture more behavior, ContractGuard extends intra-procedural acyclic paths by profiling additional virtual branches that do not exist in the code. These virtual branches check the potential pre-/post- conditions of each instruction. Specifically, for each arithmetic operation, we check the pre-condition that there is no over-/under-flow. For each function call, we check the post-condition whether the return value is null (if the return type is `address`), false (if the return type is `Boolean`), or zero (if the return type is `int`).

## 3 MOTIVATION

In this section, we present the empirical investigation results that motivate ContractGuard.

The key hypothesis in ContractGuard is that one can differentiate vulnerability-attacking transactions from normal transactions by comparing their control flow behaviors, as the formers tend to be associated with *abnormal control flow paths*, that is, paths that never occur in the training transactions (which we assume to be correct). While there is literature [12], [16] supporting this hypothesis for conventional programs, its validity for Ethereum smart contracts has yet been validated. In order to validate this hypothesis, we refer to several recent empirical studies on Ethereum smart contract vulnerabilities ([3], [6], [11], [25]–[27]). In total, we analyze 11 major vulnerabilities and investigate whether attacks on them trigger abnormal control flow paths.

**1. Reentrancy** [12]: When an innocent contract calls or sends ether to an external contract, the external calls can be hijacked by attackers, whereby forcing the contract to further execute an additional piece of code, such as calling back to the innocent contract itself. If the innocent contract does not expect reentrancy and implement the correct checking, then it is susceptible to illegal contract state changes. Attacks of this kind were used in the infamous DAO hack [13].

Depending on whether the re-entered function is the same as the function that makes the external call, the vulnerability can be classified as same-function reentrancy or cross-function reentrancy [6]. The following piece of code shows an example of same-function reentrancy. At line 6, the innocent contract makes an external call to the attacker contract. As no function name is specified, EVM will invoke the fallback function (line 12), in which `withdraw` is reentered and make illegal state changes.

```
1:   contract InnocentContract {
2:       mapping(address => uint)private balances;
3:       function withdraw(){
4:           uint amount = balances[msg.sender];
5:           if(amount > 0){
6:               msg.sender.call(balances[msg.sender]);
7:               balances[msg.sender] = 0;
8:           }
9:       }
10:  }
11:  contract AttackerContract {
12:      function (){
13:          Wallet wallet = msg.sender;
14:          wallet.withdraw()
15:      }
16:  }
```

*Abnormal control flow*: Yes. Both the same-function and the cross-function reentrancy will trigger an abnormal control flow path that reenters the innocent contract. Had the training transactions contained such control flow, the developers would already recognize the problem and fix the defect. Note that the detection requires calling-context information. This is a reason why ContractGuard profiles calling-contexts.

**2. Dangerous delegatecall** [11]: When a contract uses `delegatecall`() to call an external function, the function will be executed in the context of the calling contract. This feature enables the implementation of shared libraries. However, it could induce vulnerabilities when some library functions run in an unexpected context and make illegal state changes.

The following piece of code illustrates a simplified version of the vulnerability that cause the Parity Multisig Wallet incidence [14]. At line 11, the innocent contract uses `delegatecall`() to call a library function selected by the signature hash encoded in the message call input data. At line 15, the attacker triggers line 11 to call function `init`, which is executed in the context of `InnocentContract` and therefore transfers the ownership of `InnocentContract` to the attacker.

```
1:   contract DelegateLibrary {
2:       address public owner;
3:       function init(){
4:           owner = msg.sender;
5:       }
6:   }
7:   contract InnocentContract {
8:       address public owner;
9:       DelegateLibrary _library;
10:      function (){
11:          _library.delegatecall(msg.data);
12:      }
13:  }
14:  contract AttackerContract {
15:      … innocent.call(
16:          bytes4(keccak256("init()"))); …
```

*Abnormal control flow*: Yes. The attack of this vulnerability triggers an abnormal control flow path in which some library functions like `init` is called unexpectedly through `delegatecall`. As these library functions are never intended to be called in this way, the training transactions will never cover this abnormal control flow path.

**3. Arithmetic Over/Under Flows** [3]: In arithmetic operations, an over/under flow occurs when an operation is performed that requires a fixed size variable to store a number (e.g. `uint256`) that is outside the range of the variable's data type (e.g. `uint8`). In April 2018, malicious attackers utilized an integer overflow on multiplication to attack BECToken, causing the token price to drop to nearly zero [28].

*Abnormal control flow*: Yes. The extended paths introduced in Section 2.2 contain virtual branches that check all potential pre-conditions, and under-/over- flow absence is one of them. As we assume that the training transactions are correct, they should not cover paths that trigger under-/over-flow conditions.

**4. Default Visibilities** [26]: In Solidity contracts, the function visibility specifiers are default to `public`, allowing other contracts or users to call the functions externally. Therefore, devastating vulnerabilities may arise when the developers forget to write visibility specifiers for functions that should not be exposed to external calls.

*Abnormal control flow*: Yes. An abnormal control flow can be detected when the internal function for which developers forget to specify the visibility is called externally. The training transactions will never contain such a call, as these functions are internal in the mind of the developers from whom the training transactions are obtained.

**5. Superficial randomness** [26]: Contracts that make decision by using Ethereum state variables such as block timestamp as random seeds are subject to attacks, as these variables can all be manipulated by the miners.

*Abnormal control flow*: None. This vulnerability is essentially about misuses of variables. It is irrelevant to the control flow paths of the contracts.

**6. Unchecked send** [6]: When the contract performs a failed external call, the transaction that will normally revert. However, if the external call is made through low-level `call ()` or `send ()` API, they will return false instead of rolling back the transaction. A common pitfall arises when the return value of `call ()` or `send ()` is not checked. For example, in the following piece of code, the developers expect the transaction will revert when the external call at line 2 fails. However, when line 2 fails, the transaction will simply carry on and execute line 3, resulting in an inconsistent contract state.

```
1:  if (gameHasEnded && !prizePaidOut){
2:     winner.send(1000);
3:     PrizePaidOut = True;
4:  }
```

*Abnormal control flow*: Yes. The extended paths introduced in Section 2.2 contain virtual branches that check all potential post-conditions, and the truth-value returned by any function is one of them. As we assume the training transactions to be correct, they should not cover paths that make `call ()`/`send ()` fail.

**7. Order Dependence** [6]: It is also known as race condition [26]. The attack on this vulnerability occurs when a miner tries to "race" with a smart contract participant by inserting their own transaction before the victim's transaction in the block and making changes to the related contracts. Such changes might potentially invalidate the

*Abnormal control flow*: None. Attacks on this vulnerability target the execution environment of the contracts. There are no abnormal control flow paths to detect.

**8. Tx.Origin Authentication** [26]: Contracts that authenticate users using the `tx.origin` variable are typically vulnerable to phishing attacks which can trick users into performing authenticated actions on the vulnerable contract. For example, consider the following piece of code. The innocent contract uses the checking at line 3 to ensure that only its owner can transfer the asset. However, the attacker can trick the owner into sending ether to the attacker contract at line 8. This invokes the fallback function at line 9 and call `transfer` with `tx.origin` as the owner who is the one that initiates the transaction. Therefore, the authentication at line 3 will pass and the attacker steals the asset.

```
1:  contract InnocentContract {
2:     function transfer(address dest, uint amount){
3:        if (tx.origin != owner) throw;
4:           dest.send(amount);
5:     }
6:  }
7:
8:  contract AttackerContract {
9:     function (){
10:       Wallet w = Wallet(walletAddr)
11:       w.transfer(thiefAddr, msg.sender.balance);
12:    }
13: }
```

TABLE 1
ABNORMAL CONTROL FLOW OF VULNERABILITIES

| Vulnerability | Abnormal control flow |
|---|---|
| 1. Reentrancy | **Yes**. The path that re-enters the caller function that makes external call. |
| 2. Dangerous delegate call | **Yes**. The path that calls an unexpected library function through delegatecall |
| 3. Arithmetic Over/Under Flows | **Yes**. The extended path in which over-/under- flow condition checking virtual branch returns true. |
| 4. Default visibilities | **Yes**. The path that has an external call to the internal function for which developers forget to specify the visibility. |
| 5. Superficial randomness | **None** |
| 6. Unchecked send | **Yes**. The extended path in which the checking of the return value of `call ()` or `send ()` gets false. |
| 7. Order Dependence | **None** |
| 8. Tx.Origin Authentication | **Yes**. The path that initiates from the calling of the attacker's contract by the owner, and then calls back to the function with the `tx.origin` checking |
| 9. Denial Of Service | **Yes**. The path that leaves the contract inoperable. |
| 10. Short Address | **None** |
| 11. Logic error | **Not always exist**. Depends on whether coincidental correctness occurs in the training transactions. |

*Abnormal control flow*: Yes. In order for this attack to succeed, the attacker must trigger a transaction that initiates from the calling of the attacker's contract by the owner, and then calls back to the function with the `tx.origin` checking. By contrast, the training transaction will only call the function with `tx.origin` checking directly.

**9. Denial Of Service** [26]: This category is very broad, but fundamentally consists of attacks where users can leave the contract inoperable for a small period of time, or in some cases permanently.

*Abnormal control flow*: Yes. Apparently, the control flow path that leaves the contract inoperable is abnormal, as none of the training transactions will try to block the contracts from providing services.

**10. Short address** [26]: When the parameters passed to a smart contract are shorter than the expected parameter length, the EVM will pad 0's to the end of the encoded parameters. If a third party application does not validate inputs, it will be vulnerable to attacks that exploit this feature.

*Abnormal control flow*: None. The target of attacks on this vulnerability is not the contracts themselves, but the third party applications that interact with the contracts. Therefore, there are no abnormal control flow paths.

**11. Logic error** [6]: Complex predicates are prone to various logic-related defects such as wrong relational/logic operators and misplaced parentheses. Some of these detects can cause subtle vulnerabilities in the contracts.

*Abnormal control flow*: No always exist. If there are no overlaps between paths that trigger the failure-inducing condition of the defect and paths that do not, then abnormal control flow paths can be detected from the attacking transactions. However, it is well known that because of *coincidental correctness* [29], this requirement is not always met. Therefore, abnormal control flow paths do not always exist for attacks on this vulnerability.

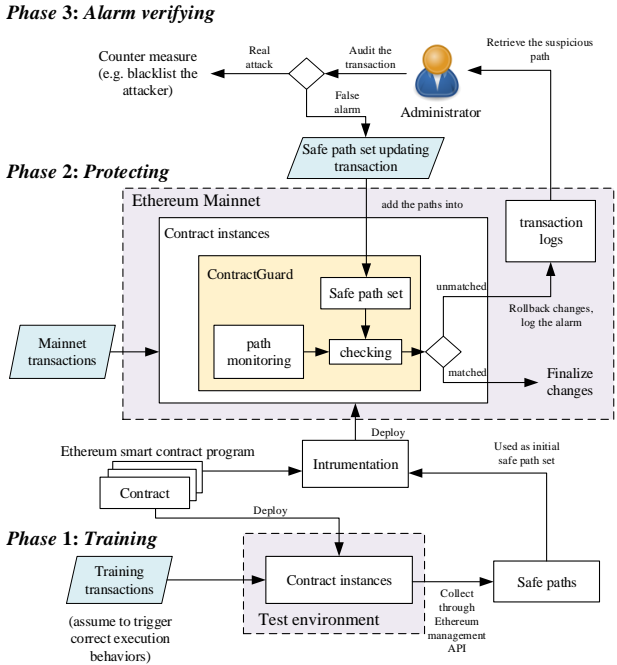

Fig. 4. The overview of ContractGuard.

TABLE 1 summarizes the result. As we can see, the attacks on a majority of the vulnerabilities are associated with abnormal control flow paths. This provides initial evidence that control flow anomaly based IDS can be effective at defending the smart contracts. For the remaining vulnerabilities, the direct attack target is not the contracts themselves: it is either the Ethereum execution environment (#5. Superficial randomness and #7. Order Dependence), or the third-party applications that interact with the contracts (#10. Short address).

# 4 THE APPROACH OF CONTRACTGUARD

## 4.1 Overview

Figure 4 shows the overall approach of ContractGuard, which consists of three phases.

In the training phase, we deploy the protected smart contract program without instrumentation on the test environment of Ethereum. There can be three choices: a private network, the Ropsten test network, or the simulator Ganache [30]. The last choice is preferred as it provides a clean environment and quick transaction processing speed. Next, we send the training transactions to the deployed contract instances. These training transactions usually come from the passed test cases provided by the developers. After executing each test case, we collect all execution traces, and then extract the covered context-tagged acyclic paths. We use these paths as the initial safe path set. As the membership checking on the initial set will be cheaper than that on the set of paths appended after deployment, we recommend developers to provide as many passed test cases as possible.

In the protecting phase, we instrument the binary code of the protected contracts with the implementation of context-tagged acyclic path profiling and the safe path set membership checking. The administrators then deploy the instrumented code on the Ethereum mainnet. After deployment, ContractGuard profiles the context-tagged acyclic paths for every transaction sent to the protected contracts and test their membership in the safe path set. If ContractGuard finds at least one abnormal path, it will roll back the state changes done by the transaction and raise an alarm.

In the alarm-verifying phase, the administrators receive alarm notification by monitoring the Ethereum mainnet transaction logs. They then review each rejected transaction by replaying it in the test environment forked from the blockchain state right before the contract receives this transaction. If this transaction is confirmed as legitimate, then the administrators will send an administration transaction to the protected contract and append the context-tagged acyclic paths covered in this transaction into the safe path set.

## 4.2 Protection Boundary of ContractGuard

Under the Ethereum smart contract program model (Section 2.1), there can be three different scenarios that the trigger IDS protection, as illustrated in Fig. 5.

In the setting of Fig. 5, the smart contract program that the administrator want to protect contains two contracts. Their deployed instances are *A* and *B*, respectively. ContractGuard instruments both of them. At the runtime, these two contract instances might interact with each other or with another unprotected contract instance *C* through message calls.

In the first scenario, the transaction starts with a call to a public/external function (referred to as the transaction entry function) of a contract within the boundary and never makes calls to unprotected contracts outside the boundary. In this scenario, ContractGuard will profile the paths for all functions in all protected contracts. When the protected contracts call each other, we insert the current calling-context information inside the message call data, so that the full calling-context can be established. Whenever a context-tagged acyclic path ends, ContractGuard checks its membership in the safe path set. If it is not found, ContractGuard will set an anomaly flag. This flag will be cascadingly passed with the message call return data toward the transaction entry function. Before the transaction entry function returns, if ContractGuard finds the anomaly flag is set, it will raise an alarm and execute RE-VERT, whose semantics dictate that all the changes made in the current transaction are rolled back [18].

In the second scenario, the transaction starts with a call to a protected contract. During execution, the transaction calls unprotected contracts, which in turn might call back to the protected contracts. For example, in Fig. 5, the entry function in A calls a function in C, which in turn calls a function in *B*. In this scenario, ContractGuard will profile the paths in both A and B. However, A and B will independently check the anomaly flag and execute REVERT when the anomaly flag is set. This results in two different outcomes depending on where the alarm is raised. If the alarm is raised in B, then only the changes made in B are rolled back. If the alarm is raised in A, then all the changes made in A, B, and C are all rolled back. We note that ideally, we shall have a consistent treatment for both cases. However, in order to achieve this, protected contracts have to explicitly synchronize the anomaly flag with each other by making extra message calls and expensive contract storage writings, because unprotected contracts are not controlled by ContractGuard. This synchronization will make the runtime overhead too high. Therefore, we make this design trade-off by not passing the anomaly flag across calls to unprotected contracts.

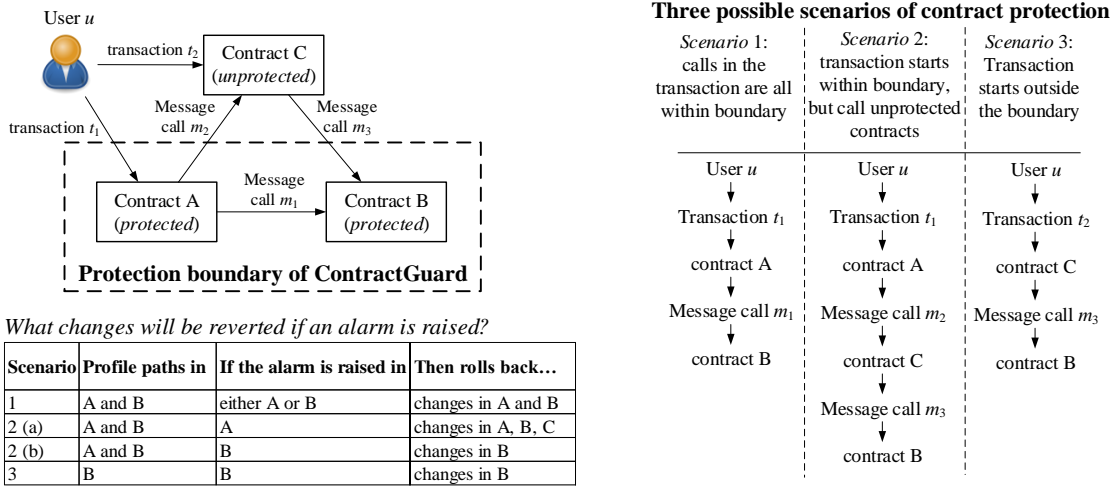

| Scenario | Profile paths in | If the alarm is raised in | Then rolls back… |
|---|---|---|---|
| 1 | A and B | either A or B | changes in A and B |
| 2 (a) | A and B | A | changes in A, B, C |
| 2 (b) | A and B | B | changes in B |
| 3 | B | B | changes in B |

Fig. 5. Protection boundary of ContractGuard and the possible protection scenarios.

For a similar reason, when a protected contract X calls an unprotected contract Z, ContractGuard does not pass the current calling-context to another protected contract Y that is transitively called by Z. One exception, though, is when X and Y are the same. In this case, ContractGuard passes the current calling-context by writing to a contract storage variable before calling the unprotected contract. We choose to implement this step because it is essential to the detection of reentrancy attacks. A drawback of this is the introduction of an extra 5000 gas per external call. As we shall see in Section 6.4, this is in fact the single most costly factor that affects the runtime overhead.

In the final scenario, the transaction starts with a call to an unprotected contract, which then calls the protected contracts. The following treatment, however, is essentially the same with those in the first two scenarios.

### 4.3 Intra-procedural Path Indexing and Profiling

ContractGuard adapts the EPP algorithm proposed by Ball and Larus [21] to index and profile intra-procedural path. It applies the adapted algorithm to each acyclic control flow graph (CFG), where backedges in the normal CFG are replaced by surrogate edges. We have introduced this treatment in Section 2.2.

Figure 6 gives the adapted EPP algorithm with an example.

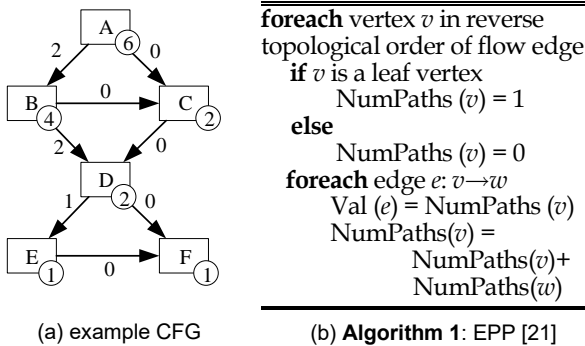

(a) example CFG     (b) **Algorithm 1**: EPP [21]

Fig. 6. Efficient intra-procedural path indexing and profiling.

The core idea is to label the acyclic CFG with two quantities. For a vertex $v$, it is labeled with NumPaths ($v$) as the number of paths from $v$ to exit vertex. For example, the vertex $D$ in Fig. 6 has two paths to the exit vertex $F$: $D \rightarrow F$ and $D \rightarrow E \rightarrow F$, therefore it is labelled with 2. Next, let $e_1, e_2 \ldots e_n$ be all the edges that starts with the same vertex $v$. Then edge $e_1$ is labelled with zero and the remaining $e_i$ is labelled with:

$$Val(e_i) = \sum_{j<i} NumPaths(end\_vertex(e_j))$$

With this labeling scheme, it can be proven (Theorem 1 in [21]) that for a vertex $v$, each path $p$ from $v$ to the exit vertex will be assigned with a unique number in the range 0, 1, … NumPaths ($v$)-1 by summing up Val($e$) for edges in $p$. Besides, it can be shown that the algorithm in Fig. 6 will generate the correct label for every vertex and edge.

After EPP labeling, we now have an efficient method to profile the acyclic intraprocedural path: At each function entry, we allocate an integer local variable `epp` with initial value 0. Next, whenever the control flow passes through an edge $e$, we add Val ($e$) to `epp` if Val ($e$) is not 0. At the function exit, we generate the covered path indexed by the value of `epp`. Besides, at the backedge, we also generate an acyclic path indexed by the value of `epp`. The variable `epp` is then reset with 0 to start a new acyclic path.

The result of the algorithm can be further optimized by rearranging the edge value assignment to minimize the number of non-zero edge values [31]. We adopt this optimization in ContractGuard.

### 4.4 Calling-context Indexing and Profiling

Besides intraprocedural paths, ContractGuard also needs calling-context information. Inspired by the EPP algorithm, we develop an efficient calling-context indexing and profiling algorithm, as shown in Fig. 7.

Our proposed algorithm works on the acyclic call graph, which is obtained by replacing the recursive function call with surrogate-edges (Section 2.2). Each callsite in the program is represented by a call edge and a return edge.

In essence, our algorithm is the "reverse" version of EPP algorithm applied on the call graph. Our algorithm also works by labeling the vertices and edges. However, instead

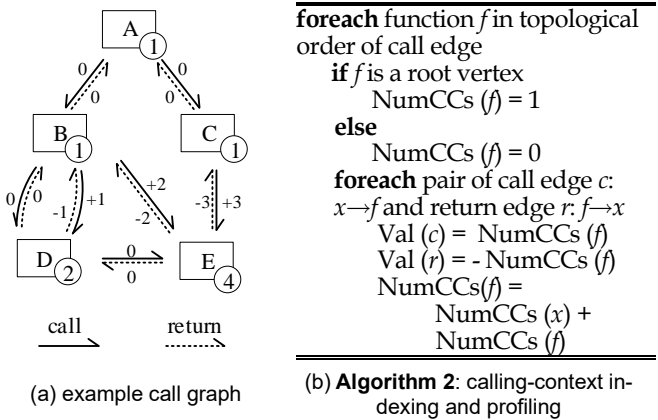

(a) example call graph    (b) **Algorithm 2**: calling-context indexing and profiling

Fig. 7. Efficient calling-context indexing and profiling.

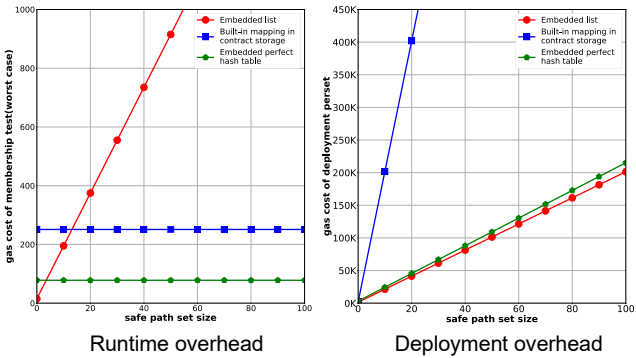

Runtime overhead    Deployment overhead

Fig. 8. Overhead trade-off of different path set storage solutions.

of following the reverse topological order, our algorithm follows the topological order. The objective is to label each vertex $f$ with NumCCs ($f$) as the number of all calling-contexts of $f$. This NumCCs ($f$) can also be interpreted as the number of call-edge paths from the program entry to $f$. Next, let $c_1, c_2 \ldots c_n$ be all the call edges that ends with the same vertex $f$, and $r_1, r_2 \ldots r_n$ are the corresponding return edges. Then edge $c_1$ is labelled with zero and the remaining $c_i$ is labelled with:

$$Val(c_i) = \sum_{j<i} NumCCs(start\_vertex(c_j))$$

, and $r_i = -c_i$ for $i$ in 1 …n.

Like EPP, we can prove that for a vertex $f$, every calling context of $f$ will be assigned with a unique number in the range 0, 1 … NumCCs ($f$)-1 by summing up Val($c$) for every call edge $c$ in the calling context. This proof is similar to that for Theorem 1 in [21]. We omit it for the sake of brevity.

To implement the calling-context profiling, we allocate a global variable ctx with initial value 0 at the program entry. At each call-site, we add Val ($e$) to ctx before the function call, then add Val ($d$) to ctx after the function return. At each function entry, the value of ctx then uniquely represents the current calling context.

### 4.5 Gas-efficient Adaptive Path Set Storage

ContractGuard combines the above two indexing schemes to create a compact indexing scheme for context-tagged acyclic path: for each function $f$ with an entry node $e$, its context-tagged acyclic paths are consecutively indexed by 0, 1, 2 … NumPaths ($e$)*NumCCs ($f$)-1. At runtime, whenever an intraprocedural acyclic path terminates, we can obtain the index of the context-tagged acyclic path as ctx*NumPaths ($e$) + epp.

With this compact indexing scheme, the problem of safe path set membership test can be formulated as follows: given an integer $k \in [0, n)$, how to check whether $k$ is in a set $S$ of $m$ integer $p_1, p_2 \ldots p_m \in [0, n)$? While this is a fundamental problem and there are many solutions, under the special gas cost model of EVM, three of them seem to make sense:

- **Embedded list**: This solution embeds $S$ into the code as a series of constant comparison.
- **Embedded minimal perfect hash table (MPHT)**: This solution embeds $S$ as a MPHT into the code. ContractGuard uses the CHD algorithm [22]. The idea of CHD is to use two levels of hashing to construct the hash table. The first level uses a hash function $G$ to group elements in $S$ into different buckets, and each buckets use a different second-level function to hash the value.
- **Built-in mapping in contract storage**: This solution uses the built-in mapping type of Solidity to store the set $S$ in the contract storage.

Figure 8 shows the performance comparison between these three solutions based on their implementation in ContractGuard. In term of deployment cost, all three solutions are linear to the size of path set. However, using built-in mapping is particularly expensive. This is because adding a path involves one SSET to write contract storage, which costs 20000 gas. By contrast, adding a path for the first two solutions involve several extra instructions, which only cost 200 gas per byte. In term of membership test gas cost, both embedded MPHT and built-in mapping require constant gas, while embedded list requires linear gas. However, embedded list costs the least gas when the set size is lower than six.

From the above analysis, we design the adaptive path set storage as follows. For the initial safe path set obtained during the training phase, ContractGuard instruments it either as an embedded list when the size is lower than six, or as an embedded MPHT otherwise. Built-in mapping is usually not used to implement the initial safe path set. However, it is always used to store the additional path appended by the administrators dynamically after deployment.

## 5 SIMULATION EXPERIMENTS ON THE MAINNET

To evaluate the efficiency of ContractGuard, we conduct simulation experiments concerning whether embedding the IDS into smart contracts is practical in term of overhead (Section 5.1), and whether there are too many false alarms for the IDS to be useful (Section 5.2).

### 5.1 Is the Overhead of ContractGuard Practical?

Protecting a smart contract program with an embedded intrusion detector in general and ContractGuard in particular involves two kinds of overhead, both of which cost extra ether cryptocurrency:

1) **Deployment overhead**: The contract code expands in size after instrumentation with the anomaly detection functionalities and the initial safe path set. Therefore, Ethereum will charge additional gas at the contract

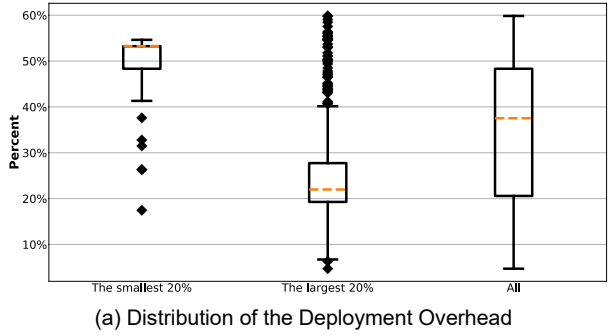

(a) Distribution of the Deployment Overhead

(b) Contracts with top-5 worst deployment overhead

| # | Contract Address on Ethereum mainnet | Original code size | Deployment overhead |
|---|---|---|---|
| 1 | 0x5bde2f1f37f594fd610b17 31284f5a204e3ea545 | 4295 | 61.23% |
| 2 | 0x80aa81029df9afdc70a621 c86d7a81d7e9ed7e3a | 5899 | 59.83% |
| 3 | 0x051fda7486480dd5abcf5 dd742ef002a2ebb9ea0 | 7703 | 59.75% |
| 4 | 0xc723d744dd32780c6f5cef 4705e99919e77879d7 | 6273 | 59.06% |
| 5 | 0x5c543e7ae0a1104f78406c 340e9c64fd9fce5170 | 2650 | 58.47% |

Fig. 9. Deployment overhead measured on all the 8,314 Ethereum mainnet contracts with at least one hundred transactions up to block number 4,200,000 (reached at the date Aug-24-2017).

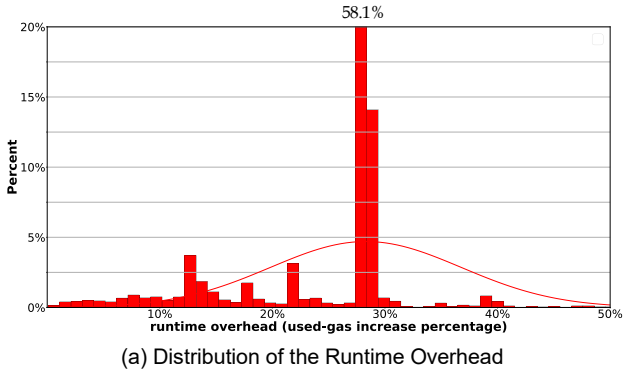

(a) Distribution of the Runtime Overhead

(b) Transactions with top-5 worst runtime overhead

| # | Transaction hash on Ethereum mainnet | Original gas /overhead |
|---|---|---|
| 1 | 0xda081509e17dcb852c4e12bc435be6d507 499993590e0da97aa0da4c1f3d915d | 22258/86.0% |
| 2 | 0xbf7a5b2113966a264dcd213d0e2a1d7d6 ce462c1f4f2059a81afe7f36be600f5 | 32529/85.7% |
| 3 | 0x832d0282d84221ad21a3e8c43652653b6 6aea8bedfafc76c357e47da2c02314b | 20022/85.4% |
| 4 | 0xd8626e496a04c8a894094eeb848cf9502d 208834fd97bbcf4c86527c10e8555f | 26693/84.5% |
| 5 | 0x41000decd3cd39be9895e3c2a7250deb5 7c2e8e99a2f59534657d71df2c428de | 43357/76.7% |

Fig. 10. Runtime overhead measured on all the 19,944,547 transactions submitted to the 8,314 Ethereum mainnet contracts.

creation, adding to the deployment cost.

2) **Runtime overhead**: During transaction execution, the embedded IDS profiles the control flow paths. Besides, when a path terminates, the IDS needs to check its membership in the path set. Both operations add to the gas cost of the transaction.

While only the administrators pay for the first overhead, all users that send transactions to the protected contracts pay for the second overhead. Had either overhead been too high, the contract administrator would not use the IDS, even if it can effectively protect their contracts.

In order to investigate the overhead of deploying Contract-Guard for contracts in the real world, we ran the latest version of Geth (1.9.0) in full archive mode (which retrains all historical information) and synchronized with Ethereum mainnet up to block number 4,200,000 (reached at the date Aug-24-2017). In total, there is 341,585 contract accounts. Among them, 332,008 contract accounts have few than 100 transactions, another 1,173 ones make the control flow graph extraction tool crashed (we use Vandal [7] in our experiments). We chose the remaining 8,314 contract accounts as the subjects for investigation, as they likely represent contracts of real use. In total, users sent 19,944,547 transactions to these contracts during the period. We compare the code passed as input data of the contract creation transactions of these 8,314 instances, and found 985 contract accounts are distinctive. The others are clones of them. In the following, we will report the average for both the complete 8,314 and the distinctive 985.

We apply ContractGuard to all the subject contract accounts. Since the cost of deploying contracts is proportional to the binary code size (200 gas per byte [18]), we measure the deployment overhead as (instrumented code size) / (original code size) - 100%. We set the safe path set of each contract contains all the context-tagged acyclic paths covered by all the transactions sent to this contract. This treatment simulates the worst case of deployment overhead.

Figure 9 (a) shows the deployment overhead for all, the largest 20%, and the smallest 20% contracts in term of their original size. The average deployment overhead is 24.07% for the largest 20% contracts and 51.84% for the smallest 20% contracts. Overall, the average deployment overhead is 36.14% for all the 8,314 contract instances and 37.76% for the 985 distinctive ones. The contracts with top-5 worst deployment overheads are shown in Fig. 9 (b).

The result in Fig. 9 provides initial evidence that the deployment overheads of IDS for smart contracts can be practical: essentially, the administrator is paying 36.14% more gas in exchange for more security for their contracts. Note that none of the 8,314 contract accounts exceeds the 24576-bytes limit after ContractGuard instruments them.

Next, we investigate the runtime overhead. For all the 19,944,547 transactions sent to the subject contracts, we collect their detailed execution traces. We then align each trace with the code instrumented by ContractGuard, and make up the instructions that the transaction would have executed had ContractGuard been embedded into the contract. In this way, we obtain the precise runtime overhead of ContractGuard on each transaction by accumulating the extra gas cost spent on executing the extra instructions.

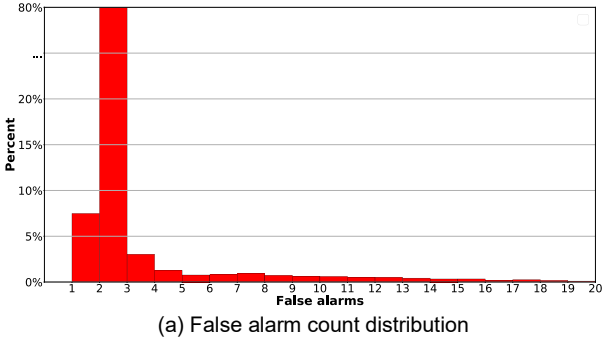

(a) False alarm count distribution

| # | Contract Address on Ethereum mainnet | False alarm count |
|---|---|---|
| 1 | 0x49516fe7bdc54b29a7f95ff55fdd38d9228e55af | 60 |
| 2 | 0x2faa316fc4624ec39adc2ef7b5301124cfb68777 | 59 |
| 3 | 0xda4a4626d3e16e094de3225a751aab7128e96526 | 56 |
| 4 | 0x3f593a15eb60672687c32492b62ed3e10e149ec6 | 53 |
| 5 | 0xb2c818252457488728ca62affa8f320247b8a84a | 53 |

(b) Contracts with top-5 most false alarm counts

Fig. 11. Upper-bound Estimation of the False Alarm Counts.

Figure 10 (a) shows the distribution of the runtime gas overhead in terms of gas increase. On average, the runtime overhead is 25.98% for the 8,314 instances and 17.5% for the 985 distinctive ones. This is comparable to that of practical control flow anomaly based IDSs for conventional systems (e.g. the average execution time overhead of the work by Giffin and colleagues is 22% [19]). For a majority of the transactions (94.5%), the runtime gas overhead is under 30%.

The transactions with top-5 worst runtime overheads are shown in Fig. 10 (b). In the worst case, the runtime gas overhead is 86.0%. We have performed a detailed analysis on the factors that affect the runtime overhead (presented in Section 6.4). We found that the reason why the runtime gas overhead is so high for a small amount of the transactions is that they made an exceptionally high number of calls to external unprotected contracts. When this happens, ContractGuard has to use the expensive contract storage to persist the calling contexts. At the moment, ContractGuard might be inapplicable to such contracts (under 5% among the 8,314 subject contracts). We present more discussion on this issue in Section 6.5.

### 5.2 Is It Practical to Handle False Alarms Manually?

Besides overheads, another important issue that might affect the practical use of ContractGuard is false alarms [32], which correspond to legitimate transactions that have been mistakenly classified as malicious. To avoid rejecting future legitimate transactions, the administrators need to review each alarm manually and append the path into the safe path set if it is a false alarm. This potentially involves significant manual effort.

In order to investigate the amount of false alarms. We use the 8,314 real world contract accounts with at least one hundred transactions. We assume that the developers deploy their contracts without training. Therefore, ContractGuard will have an empty safe path set initially. We further assume that all the transactions are legitimate, and the administrators will immediately add the context-tagged acyclic paths of a transaction to the safe path set when this transaction is reported as a false alarm. Therefore, ContractGuard will not report the following transactions with the same context-tagged acyclic paths as false alarms. Under these assumptions, we then count the amount of false alarms for each contract. Essentially, this is an upper-bound estimation of the manual effort required to handle false alarms in the real world.

Figure 11 (a) shows the distribution of false alarms, while Fig. 11 (b) shows contracts with top-5 most false alarms. Under the hypothesized scenario, ContractGuard reports 2.88 false alarms on average for each contract. For a majority of the transactions (92.2%), the number of false alarms is under 5. This result lends support to the conjecture that the false alarm issue is manageable when we apply IDS to defend contracts.

## 6 CONTROLLED EXPERIMENTS

We present the experimental evaluation of the effectiveness of ContractGuard in this section. Three research questions will be evaluated:

**RQ1**: Can ContractGuard effectively detect vulnerabilities?

**RQ2**: What are the deployment overhead, runtime overhead, and the amount of false alarms?

**RQ3**: What are the factors that affect the deployment and runtime overheads?

This section describes the experiment setup to evaluate our research questions and reports the experimental results.

### 6.1 Experimental Subjects

To investigate RQ1 and RQ2, we design two sets of experiments. The first set of experiments (TABLE 2) uses six real-life contract programs with reported vulnerabilities and attack vectors. We choose these subjects because they represent famous attacks that have caused significant damage. For example, the first subject "DAO" is the original contract from which hackers managed to steal tokens valued $60M and ultimately led to the hard-fork that created Ethereum Classic (ETC) [6]. And the second subject "Multisig" contains an exploit [7] that ultimately causes about $300M-worth cryptocurrency to be frozen and (probably) lost forever. The fourth contract subject "BecToken" is the one that was attacked by malicious attackers in April 2018, causing the token price to drop to nearly zero [28].

The second set of experiments (TABLE 3) uses real-life contract programs whose projects are open source and contain test cases written by developers in the Truffle framework. We choose these subjects because they are some of the most complex dapp hosted on Github in term of code size. We publish all the subjects, test cases, vulnerabilities, and their attack vectors on our project website (https://github.com/contract-guard/experiments).

To investigate RQ3, we use the 8,314 contract instances from the Ethereum mainnet. Explanation of the experimental subjects can be found in Section 5.1.

### 6.2 Empirical Study with Real-life Vulnerabilities

For each of the six smart contract programs with a reported vulnerability and an attack vector (TABLE 2), we strictly follow the report to construct the attacking transaction. This attacking transaction is then mixed with 99 non-reverting normal transactions randomly chosen from the

TABLE 2
EXPERIMENTS WITH REAL-LIFE REPORTED VULNERABILITIES AND ATTACK VECTORS

| Program | Vulnerability and reported attack vector | Code Size (bytes) | Deployment Overhead % | Deployment Overhead avg. gas | Runtime Overhead % | Runtime Overhead avg. gas | Recall avg. | False alarms avg. |
|---|---|---|---|---|---|---|---|---|
| DAO | Reentrancy [33] | 1774 | 11.95% | 35751.43 | 20.51% | 43845.47 | 100% | 4.48 |
| MultiSig | Dangerous Delegatecall [34] | 764 | 29.58% | 22679.66 | 28.27% | 29085.6 | 100% | 2.0 |
| KoETH | Unchecked send [35] | 3617 | 31.18% | 37359.64 | 23.82% | 44067.72 | 100% | 13.57 |
| BecToken | Arithmetic overflow [36] | 5963 | 9.52% | 59846.26 | 15.75% | 68371.95 | 100% | 8.2 |
| OwlWallet | Tx.Origin Authentication [37] | 561 | 27.27% | 25032.01 | 26.50% | 31456.74 | 100% | 3.0 |
| MerdeToken | Arithmetic underflow [38] | 1013 | 21.03% | 35990.97 | 18.74% | 42397.30 | 100% | 4.0 |

TABLE 3
EXPERIMENTS WITH REAL-LIFE DEVELOPER TEST CASES AND SEEDED VULNERABILITIES

| Program and project site | Code Size (bytes) | Truffle test cases[†] | Total seeded Vulnerabilities | Subject seeded Vulnerabilities | Deployment overhead % | Runtime overhead % | Recall avg. | False alarms avg. |
|---|---|---|---|---|---|---|---|---|
| Cryptokitties [40] | 30966 | 516 | 64(14,7,3,40) | 5(2,1,0,2) | 14.28% | 9.03% | 60.0% | 1.5 |
| StatusNetwork [41] | 38631 | 261 | 83(28,9,6,40) | 2(0,1,1,0) | 19.02% | 15.79% | 100% | 3.0 |
| VotingDapp [42] | 2326 | 14 | 15(6,2,0,7) | 2(2,0,0,0) | 12.33% | 10.84% | 100% | 1.0 |
| pool-shark [43] | 5895 | 69 | 33(8,6,1,18) | 3(2,1,0,0) | 10.23% | 31.52% | 100% | 5.9 |

[†] *The number of test cases is measured by the total number of transactions sent to the contracts when executing the complete Truffle test suite.*

Ethereum mainnet to create a test case as a sequence of 100 transactions. The attacking transaction is put randomly at the last 10% positions of the sequence. To avoid bias, we repeat this process to create 100 test cases with different random normal transactions from the mainnet. The contract creation transactions are separately achieved from the mainnet and shared among test cases.

As we do not have the developer test suites for these programs, we omit training and start with an empty safe path set. We assume that all false alarms raised at the normal transactions will be examined by the administrators, and these transactions' paths will be inserted into the safe path set immediately. Under this assumption, we apply ContractGuard to each program and execute them with the 100 test cases, and measure *recall* as the percentage of test cases in which Contract successfully raises alarms at the attacking transactions.

TABLE 2 shows the experiment results. For all of the programs, the recall is 100%. This shows how effectively ContractGuard can defend against these known vulnerabilities. On the other hand, the runtime overhead, the deployment overhead, and false alarms are consistent with those observed on the 8,314 contract accounts on the mainnet (Section 5.1). As of August 30 2019, the average gas price is $1.4683 \times 10^{-8}$ ETH, while the ETH price is $168.21 USD [39]. As shown in TABLE 2, the maximum deployment and runtime overhead are 59846.26 gas and 68371.95 gas, respectively. Therefore, if the administrators apply ContractGuard on these six real world contracts, at the most they need to pay an extra $0.1478 USD for each contract at deployment. And at the most the users need to pay an extra $0.1689USD for each transaction.

### 6.3 Empirical Study with Seeded Vulnerabilities

The programs in TABLE 3 come with open source project sites and test suites developed in the Truffle framework. These test suites are ideal candidates for training when we deploy ContractGuard to real-world contracts. These programs, though, do not have any reported vulnerabilities. To facilitate the experiment, we refer to the eleven major vulnerabilities in TABLE 1 and seed four of them:

- #3 **arithmetic over-/under- flow**: replacing Safe-Math [44] calls with direct arithmetic operators.
- #4 **default visibility**: removing keyword "private"
- #6 **unchecked send**: removing the whole branch that check the return value of `send` and `transfer`.
- #11 **logic error**: replacing relational and logic operators.

Seeding the remaining seven types of vulnerabilities is difficult either because they are not directly related to the contract code (#5, #7, #10), or because these programs do not have the related program structures that enable seeding (#1, #2, #8, #9). Note that #1, #2, #8 have already been covered in the first experiment.

For each program, we create all possible seeded versions. The total number and those for the four vulnerabilities are shown in the "total seeded" column of TABLE 3. We then execute the developer test suite on each seeded version. Versions that fail the test suite are discarded. Next, we create a test case by randomly draw 100 transactions from both the mainnet and a test pool we manually construct to trigger the attack on the seeded vulnerability, and then execute the test case with both the original program and the seeded version. If their contract storage values are different after executing a transaction, then we add this version into the subject seeded version set and mark this transaction as the attacking transaction. This process is repeated until we have 100 test cases for each subject seeded version. We use these subject seeded versions to simulate vulnerabilities that the developer test case fail to detect, but can be exploited by hackers after deployment. Their total number and those for the four vulnerabilities are reported in the "subject seeded" column of TABLE 3.

For each subject seeded version, we apply ContractGuard on them and use the developer test suite for training. We then execute the 100 test cases for this version, and measure the *recall* of this version as the percentage of test cases in which Contract successfully raises alarms and reverts at the simulated attacking transactions. As there are multiple seeded versions, we report the average recall for each program.

TABLE 3 shows the experiment results. The average recall is 83.3% for all subject seeded versions of all programs. We examine the missed cases and find that all of the versions that ContractGuard misses are seeded with logic errors. The reason why ContractGuard fails to prevent them is because of coincidental correctness: there exists some developer test cases which execute the same path as the attacking transaction, but did not trigger the failure. As the paths of the attacking transactions are in the safe path set, ContractGuard fails to detect them.

TABLE 4
OVERHEAD OF VARIOUS INSTRUMENT POINTS (IPs)

| Instrument point | Deployment gas cost | Runtime gas cost |
|---|---|---|
| Contract wrapper | 16000 | 500~5980 |
| Function entry | 4200 | 45 |
| Branch | 2600 | 30 |
| Backedge | 5200 | 57 |
| Internal call-edge | 4400 | 48 |
| Internal return-edge | 6200 | 68 |
| External call to unprotected contract | 5000 | 5348 |
| External call to protected contract | 6800 | 70 |
| Function exit | 800 | 9 |
| Path set checking[†] | 1600*$n$ + 2800 | 378 |

[†]: *n is the size of the initial safe path set*

### 6.4 Factors that Affect Overheads

In order to investigate the factors that might affect the overhead, we decompose the implementation of Contract-Guard into various *instrument points* (IP), each of which represents a point in the original code to insert instrumented code. Due to the lack of space, we leave the detailed implementation in the supplementary material.

TABLE 4 shows the deployment/runtime gas cost of each individual instrument point. The deployment cost is simply the byte size of instrumented code multiplied by 200. The highest cost comes from the contract wrapper instrumented at the beginning of every contract. This is the reason why smaller contacts tend to have higher deployment overhead.

The runtime cost is calculated with the gas model of EVM [18]. As we can see, besides the upfront cost of contract wrapper, the most costly factor is the external call to unprotected contract. The high cost is due to the fact that we need to persist `ctx` into the contract storage: a single `SSET` instruction that rewrite a 256-bit word in the storage already cost 5000 gas. This is the trade-off we need to make in order to detect the reentrant control flow paths. Note that external call to protected contract is significantly cheaper, because in this case `ctx` can be passed inside the message call data without the need of expensive writing to contract storage.

### 6.5 Discussions

We now discuss several issues that arise in the experiments.

1. At the moment, ContractGuard might not be very effective at defending against logic error in a complex predicate due to coincidental correctness. As one of the main causes of coincidental correctness is short-circuiting, we conjecture that by adding virtual branches that evaluate every short-circuited condition, the effectiveness of ContractGuard can be improved. We plan to explore more on this issue in the future.

2. The high cost of external call to unprotected contracts prevent the practical application of ContractGuard to smart contract programs with an abundance of this feature. While this seems a limitation, we shall note that making an external call to unfamiliar contracts is a dangerous practice and highly discouraged [26]. On the other hand, ContractGuard does not introduce high cost when the contract makes an external call to protected contracts within the same program, which is a common practice in multi-contracts programs.

3. Although we haven't found any such cases among existing Ethereum mainnet contracts, it is still possible that the size of a contract exceeds the 24576 bytes hard limit when it is instrumented by ContractGuard. When this happens, the most probable cause might be that the initial safe path set is too large to be embedded into the code as a list or MPHT. In this case, ContractGuard will resort to using built-in mapping in contract storage to store some of the paths. This will significantly increase the deployment overhead, as the administrators need to pay for extra expensive transactions to write into the built-in mapping (20000 gas for each path). However, this will not increase the runtime overhead, because each set membership test always contain a visit to the built-in mapping for checking against dynamically added paths, and this visit costs constant gas regardless of the size of the mapping.

## 7 RELATED WORK

### 7.1 Smart Contract Analysis and Verification

In the literature, a number of tools have been proposed to verify smart contracts. They can be classified according to the underlying techniques. Some of them rely on program analysis to find vulnerabilities. For example, tools including Oyente [45], Teether [4], Gasper [5] and the recent work by Grossman et al. [46] use symbolic execution to explore whether there exists paths that can trigger any known vulnerability. ContractFuzzer [11] uses random fuzzing to find potential attack vectors. Other works rely on formal verification and theorem proving. For example, Zeus [6] uses abstract interpretation and constrained horn clauses, MadMax [10] uses Datalog theorem prover, and the work by Grishchenko et al. [47] uses the F* theorem prover. Grishchenko et al. [47] uses the F* theorem prover.

Although these tools can detect candidates of vulnerabilities, the use of them alone may not adequately protect the execution of smart contract programs. Most of these tools are designed to deal with specific *known* vulnerabilities. Their protection against *unknown* vulnerabilities is uncertain. In contrast, vulnerabilities detected by ContractGuard based on control flow anomaly is not limited to a specific known pattern. It can defend against unknown vulnerabilities as long as they trigger a control flow path not observed in training. Besides, as a tool for use *after* deployment, ContractGuard complements existing verification tools for use *before* deployment.

### 7.2 Execution Profiling

Control flow profiling is a fundamental dynamic analysis technique with many applications. Since the seminal EPP algorithm was proposed [21], many works have tried to improve EPP either by adding more control flow information, or by further reducing the overhead. For example, whole program path (WPP) profiling [48] compresses the full sequence of intra-procedural acyclic paths to profile the complete control flow path. Interprocedural path profiling [49] tries to achieve the same goal by extending EPP to the system control flow graph. Both works are too expensive to apply to Ethereum smart contracts in term of runtime overhead. The recent work by D'Elia and Demetrescu [50] improves EPP by extending the acyclic path segments across multiple loop iterations. Although the runtime overhead is comparable to EPP, this improvement might dramatically increases the path numbers for functions with nested loops. This is particularly undesirable in our scenario, as the path set size will inflate, adding to the deployment cost. Besides path profiling, there are also research on calling context profiling. For example, a recent work by Sumner and colleagues [51] utilize stack depth to optimize the profiling cost. The stack depth information, however, is not available in EVM.

Besides control flows, other aspects of execution can be profiled, such as data flows [52] or dynamic invariants [53]. The study of their usefulness to protect smart contract programs is left as future work.

### 7.3 Intrusion Detection System (IDS)

Conventional IDSs can be classified as signature-based or anomaly-based [19]. Signature-based IDSs (e.g. [54], [55]) try to identify known patterns of intrusions with pre-identified intrusion signatures, while anomaly-based IDSs (e.g. [15], [16], [19]) assumes the nature of the intrusion is unknown, but will deviate from the program's normal behaviors profiled from training. Signature-based IDSs in general are more precise, as they can utilize vulnerability-specific properties to detect attacks. However, they only work on known vulnerabilities. Furthermore, though not a major concern for conventional systems, the signature database might be too large for IDS to store them in the contracts. These are the reasons why ContractGuard follows the anomaly-based approach instead.

Existing anomaly-based IDSs use different models to define the behavior of the protected systems. There are two main dimensions to the model. The first dimension is the profile execution information. Examples include system calls [19], call stacks [16], input data/network packets [19], and jump sequence [16]. The second dimension is how the execution information is abstracted into a model. Besides the simple sequence model, other more sophisticated models have been proposed, such as the Dyck model [19], the *n*-jump sliding window model [15], and the waypoints model [16]. ContractGuard adds to this line of research with the new context-tagged acyclic path model designed to meet the stringent requirements of protecting Ethereum smart contract programs.

## 8 CONCLUSIONS

From the perspective of administrators, Ethereum smart contract programs are vulnerable for attacks after deployment. Once a contract has been deployed, anyone from anywhere can anonymously inspect its implementation and current contract states, at any time. Once a vulnerability is found, the hackers can immediately make an intrusion attack as long as they have enough ethers. The damage thus made will be irreversible once it has been accepted into the main chain. *There is no further chance to fix the mistake*!

In this work, our goal is to address this undesirable situation and provide a second chance to the administrators. We propose ContractGuard, which is the first IDS for Ethereum smart contracts. By applying ContractGuard on real world Ethereum mainnet contracts, real world vulnerabilities, and seeded vulnerabilities, we show that ContractGuard is effective at defending vulnerabilities with low deployment and runtime overheads.

## 9 ACKNOWLEDGMENT

This work was supported in part by HKSAR RGC/GRF 16202917, National Key R&D Program of China 2018YFB1404402, the Key R&D Program for Guangdong (2019B010137003), the Science and Technology Planning Project of Guangdong (2016B030305006, 2018A07071702, 201804010314, 2012224-12).

## REFERENCES

[1] S. Nakamoto, "Bitcoin: A peer-to-peer electronic cash system," *bitcoin.org*, 2008.

[2] V. Buterin, "A next-generation smart contract and decentralized application platform," *white paper*, 2014.

[3] N. Atzei, M. Bartoletti, and T. Cimoli, "A survey of attacks on ethereum smart contracts (sok)," in *International Conference on Principles of Security and Trust*, Springer, 2017, pp. 164–186.

[4] J. Krupp and C. Rossow, "teether: Gnawing at ethereum to automatically exploit smart contracts," in *27th USENIX Security Symposium (USENIX Security 18)*, 2018, pp. 1317–1333.

[5] T. Chen, X. Li, X. Luo, and X. Zhang, "Under-optimized smart contracts devour your money," in *Software Analysis, Evolution and Reengineering (SANER), 2017 IEEE 24th International Conference on*, 2017, pp. 442–446.

[6] S. Kalra, S. Goel, M. Dhawan, and S. Sharma, "ZEUS: Analyzing Safety of Smart Contracts," in *Proceedings of NDSS*, 2018.

[7] L. Brent *et al.*, "Vandal: A scalable security analysis framework for smart contracts," *arXiv preprint arXiv:1809.03981*, 2018.

[8] J. Zakrzewski, "Towards Verification of Ethereum Smart Contracts: A Formalization of Core of Solidity," in *Working Conference on Verified Software: Theories, Tools, and Experiments*, 2018, pp. 229-247.

[9] I. Nikolić, A. Kolluri, I. Sergey, P. Saxena, and A. Hobor, "Finding the greedy, prodigal, and suicidal contracts at scale," in *Proceedings of the 34th Annual Computer Security Applications Conference*, 2018, pp. 653–663.

[10] N. Grech, M. Kong, A. Jurisevic, L. Brent, B. Scholz, and Y. Smaragdakis, "MadMax: Surviving Out-of-gas Conditions in Ethereum Smart Contracts," in *Proceedings of the ACM on Programming Languages*, 2018, vol. 2, pp. 116:1–116:27.

[11] B. Jiang, Y. Liu, and W. K. Chan, "ContractFuzzer: Fuzzing Smart Contracts for Vulnerability Detection," in *Proceedings of the 33rd ACM/IEEE International Conference on Automated Software Engineering*, New York, NY, USA, 2018, pp. 259–269.

[12] H.-J. Liao, C.-H. R. Lin, Y.-C. Lin, and K.-Y. Tung, "Intrusion detection system: A comprehensive review," *Journal of Network and Computer Applications*, vol. 36, no. 1, pp. 16–24, 2013.

[13] "Understanding the DAO attack." www.coindesk.com/understanding-dao-hack-journalists.

[14] "Parity Multisig Hacked. Again." medium.com/chain-cloud-company-blog/parity-multisig-hack-again-b46771eaa838.

[15] T. Zhang, X. Zhuang, S. Pande, and W. Lee, "Anomalous path detection with hardware support," in *Proceedings of the 2005 international conference on Compilers, architectures and synthesis for embedded systems*, 2005, pp. 43–54.

[16] H. H. Feng, O. M. Kolesnikov, P. Fogla, W. Lee, and W. Gong, "Anomaly detection using call stack information," in *Proceedings of the 2003 Symposium on Security and Privacy*, pp. 62–75.

[17] T. Garfinkel and M. Rosenblum, "A Virtual Machine Introspection Based Architecture for Intrusion Detection.," in *Proceedings of NDSS*, 2003, vol. 3, pp. 191–206.

[18] G. Wood, *Ethereum yellow paper*. website, 2014.

[19] J. T. Giffin, S. Jha, and B. P. Miller, "Efficient Context-Sensitive Intrusion Detection," in *Proceedings of NDSS*, 2004.

[20] H. Xu, W. Du, and S. J. Chapin, "Context sensitive anomaly monitoring of process control flow to detect mimicry attacks and impossible paths," in *International Workshop on Recent Advances in Intrusion Detection*, 2004, pp. 21–38.

[21] T. Ball and J. R. Larus, "Efficient path profiling," in *Proceedings of the 29th annual ACM/IEEE international symposium on Microarchitecture*, 1996, pp. 46–57.

[22] D. Belazzougui, F. C. Botelho, and M. Dietzfelbinger, "Hash, displace, and compress," in *European Symposium on Algorithms*, 2009, pp. 682–693.

[23] C. Dannen, *Introducing Ethereum and Solidity*. Springer, 2017.

[24] X. Zhuang, M. J. Serrano, H. W. Cain, and J.-D. Choi, "Accurate, efficient, and adaptive calling context profiling," in *ACM Sigplan Notices*, 2006, vol. 41, pp. 263–271.

[25] X. Li, P. Jiang, T. Chen, X. Luo, and Q. Wen, "A survey on the security of blockchain systems," *Future Generation Computer Systems*, 2017.

[26] D. A. Manning, "Solidity Security: Comprehensive list of known attack vectors and common anti-patterns," *Sigma Prime*. .

[27] D. N. | T. Derivatives, "A Survey of Solidity Security Vulnerability," *duo-network*, 12-May-2018. .

[28] "BEC Spiked 4000% On First Trading Day, Another Pump-and-Dump Scheme? https://news.8btc.com/bec-spiked-4000-on-first-trading-day-another-pump-and-dump-scheme." .

[29] X. Wang, S.-C. Cheung, W. K. Chan, and Z. Zhang, "Taming coincidental correctness: Coverage refinement with context patterns," in *Proceedings of ICSE*, 2009, pp. 45–55.

[30] "Truffle Suite | Sweet Tools for Smart Contracts." [Online]. Available: https://truffleframework.com/.

[31] T. Ball and J. R. Larus, "Optimally profiling and tracing programs," *ACM Transactions on Programming Languages and Systems (TOPLAS)*, vol. 16, no. 4, pp. 1319–1360, 1994.

[32] G. C. Tjhai, M. Papadaki, S. M. Furnell, and N. L. Clarke, "Investigating the problem of IDS false alarms: An experimental study using Snort," in *IFIP International Information Security Conference*, 2008, pp. 253–267.

[33] "Understanding The DAO Attack." www.coindesk.com/understanding-dao-hack-journalists.

[34] "An In-Depth Look at the Parity Multisig Bug." http://hackingdistributed.com/2017/07/22/deep-dive-parity-bug/.

[35] "Post-Mortem." www.kingoftheether.com/postmortem.html.

[36] "Detecting Integer Arithmetic Bugs in Ethereum Smart Contracts." https://media.consensys.net.

[37] "Smart Contract Wallets created in frontier are vulnerable to phishing attacks." https://blog.ethereum.org/2016/06/24/security-alert-smart-contract-wallets-created-in-frontier-are-vulnerable-to-phishing-attacks/.

[38] *MerdeToken:* https://github.com/Arachnid/uscc/tree/master/submissions-2017/doughoyte. .

[39] *etherscan:* https://etherscan.io/chart/gasprice. .

[40] *CryptoKitty:* https://github.com/mybios/cryptocat. .

[41] *StatusNetwork:* https://github.com/status-im/status-network-token. .

[42] *DecentralizedVoting:* https://github.com/tko22/eth-voting-dapp. .

[43] *PoolShark:* https://github.com/joze144/pool-shark. .

[44] "SafeMath ·OpenZeppelin." [Online]. Available: https://openzeppelin.org/api/index.html. [Accessed: 22-Mar-2019].

[45] L. Luu, D.-H. Chu, H. Olickel, P. Saxena, and A. Hobor, "Making smart contracts smarter," in *Proceedings of the 2016 ACM SIGSAC Conference on Computer and Communications Security*, 2016, pp. 254–269.

[46] S. Grossman *et al.*, "Online detection of effectively callback free objects with applications to smart contracts," *Proceedings of the ACM on Programming Languages*, vol. 2, no. POPL, p. 48, 2017.

[47] I. Grishchenko, M. Maffei, and C. Schneidewind, "A Semantic Framework for the Security Analysis of Ethereum smart contracts," in *International Conference on Principles of Security and Trust*, 2018, pp. 243–269.

[48] J. R. Larus, "Whole program paths," in *Proceedings of the ACM SIGPLAN 1999 conference on Programming language design and implementation (PLDI)*, 1999, vol. 34, pp. 259–269.

[49] D. Melski and T. Reps, "Interprocedural path profiling," in *International Conference on Compiler Construction*, 1999, pp. 47–62.

[50] D. C. D'Elia and C. Demetrescu, "Ball-larus path profiling across multiple loop iterations," in *ACM SIGPLAN Notices*, 2013, vol. 48, pp. 373–390.

[51] W. N. Sumner, Y. Zheng, D. Weeratunge, and X. Zhang, "Precise calling context encoding," *IEEE Transactions on Software Engineering*, vol. 38, no. 5, pp. 1160–1177, 2012.

[52] G. Ammons and J. R. Larus, "Improving data-flow analysis with path profiles," in *ACM SIGPLAN Notices*, 1998, vol. 33, pp. 72–84.

[53] M. D. Ernst, J. Cockrell, W. G. Griswold, and D. Notkin, "Dynamically discovering likely program invariants to support program evolution," *IEEE Transactions on Software Engineering*, vol. 27, no. 2, pp. 99–123, 2001.

[54] V. Kumar and O. P. Sangwan, "Signature based intrusion detection system using SNORT," *International Journal of Computer Applications & Information Technology*, vol. 1, no. 3, pp. 35–41, 2012.

[55] W. Cui, M. Peinado, H. J. Wang, and M. E. Locasto, "Shieldgen: Automatic data patch generation for unknown vulnerabilities with informed probing," in *2007 IEEE Symposium on Security and Privacy (SP'07)*, 2007, pp. 252–266.

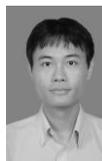

**Xinming Wang** received the Ph.D degree from the Hong Kong University of Science and Technology. He is currently a research fellow in the Lakala Group, China. His research interests include finance technology, software testing and analysis, and blockchain technology. He is a member of the IEEE.

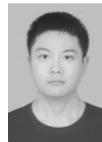

**Jiahao He** received the bachelor degree from South China Normal University. He is currently is an MSc candidate student in Computer Science. His research interests include blockchain and program analysis. He is a student member of the IEEE.

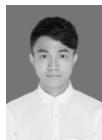

**Zhijian Xie** is currently an MSc candidate student in Computer Science. His research interests include blockchain and program analysis. He is a student member of the IEEE.

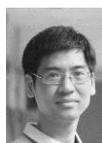

**Gansen Zhao** received the Ph.D degree from the University of Kent. He is currently a faculty member in the School of Computer Science Department, South China Normal University. His research interests include cloud computing, security, privacy, big data, and blockchain technology. He served as the program chair of the first IEEE International Conference on Cloud Computing (CloudCom2009).

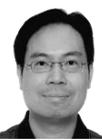

**S.C. Cheung** received his doctoral degree in Computing from the Imperial College London. He joined the Hong Kong University of Science and Technology (HKUST) where he is a professor of Computer Science and Engineering. He founded the CASTLE research group at HKUST. He was the General Chair of the 22nd ACM SIGSOFT International Symposium on the Foundations of Software Engineering (FSE 2014). He was an editorial board member of the IEEE Transactions on Software Engineering (TSE, 2006-9). His research interests focus on the quality enhancement of software for mobile, web, deep learning, open-source and end-user applications. He is a senior member of IEEE and a distinguished member of the ACM.